\documentclass[10pt,prc,aps]{revtex4}
\draft

\usepackage{graphicx}
\usepackage{subfigure}
\usepackage{color}
\usepackage{mathtools}
\begin{document}

\title{ General features of the stellar matter equation of state from microscopic theory, new maximum-mass constraints, and causality}  
\author{            
Francesca Sammarruca\footnote{Corresponding author. Email: fsammarr@uidaho.edu}  and Tomiwa Ajagbonna}                                                         
\affiliation{ Physics Department, University of Idaho, Moscow, ID 83844-0903, U.S.A. 
}

\date{\today} 

\begin{abstract}
The profile of a neutron star probes a very large range of densities, from the density of iron up to several times the density of saturated nuclear matter, and thus no theory of hadrons can be considered reliable if extended to those regions. We emphasize the importance of taking contemporary {\it ab initio} theories of nuclear and neutron matter as the baseline for any extension method, which will unavoidably involve some degree of phenomenology. We discuss how microscopic theory, on the one end, with causality and maximum-mass constraints, on the other, set strong boundaries to the high-density equation of state. We present our latest neutron star predictions where we combine polytropic extensions and parametrizations guided by speed of sound considerations. The predictions we show include our baseline neutron star cooling curves.

\end{abstract}
\maketitle

\section{Introduction} 
\label{Intro} 
A fully microscopic equation of state (EoS) up to central densities of the most massive stars -- potentially
involving phase transitions and non-nucleonic degrees of freedom -- is not within reach.
Nevertheless, neutron stars are powerful natural laboratories for constraining theories of
the EoS~\cite{Abb17a, Abb17b, Abb18, Abb19, Mil19, Mil21}. One must be mindful of the theory's limitations and the best ways to extract and interpret
 information from observational constraints. Recently, detection of gravitational waves from merging of binary neutron star systems provided constraints on both their radius and tidal deformability.

Large Bayesian interference analyses have become popular as a tool to constrain the properties of neutron-rich matter.  An example is Ref.~\cite{Huth+22}, where the authors sample 15,000 EoSs, together with observational constraints and heavy ion collision (HIC) data.  These analyses are important, but one must be careful about interpretation -- relating HIC observables to parametrizations of the EoS is not a model-independent process. It is therefore not surprising that the authors of Ref.~\cite{Huth+22} find that the HIC constraints
tend to prefer stiffer EOSs than those favored by astrophysical observations, and, we add, stiffer than those generated by {\it ab initio} theory. The reasons can be found in the phenomenological density functionals inspired by Quantum Hydrodynamics (QHD), often used to relate HIC observables to the EoS parameters. This point will be discussed in sect.~\ref{high}. Recent empirical determinations of the stellar matter EoS from data and observations have been reported~\cite{Tsang+2024, Davis+2024}.
 A Bayesian inference study aimed at assessing the performance of the Skyrme energy density functionals can be found in Ref.~\cite{Klaus+2025}. 

When using sophisticated statistical techniques, it's important not to lose sight of basic physics arguments, such as the importance of a realistic description of few-body data. An extensive discussion on this point can be found in Ref.~\cite{Sam24}.

 In this paper, keeping a firm foot in the microscopic theory -- that is, with no adjustments of nuclear forces in the medium -- we wish to illustrate {\it general features} of the EoS in different density regions, based only on theory (for normal to moderately-above-normal densities), and a few robust constraints, such as causality and the most recent maximum-mass constraints~\cite{mmax} (for high and superhigh densities). 

The cooling properties of neutron stars, observationally
accessible in terms of temperature (or luminosity) {\it vs}. age
relations, are also an important tool to obtain a glimpse on the internal
structure and composition of these exotic systems.
Ages and thermal luminosities of neutron stars, inferred from observations, can be interpreted with the aid of the neutron star
cooling theory to gain information on the properties of superdense matter in the interior of the star.
We present our first results of cooling simulations, and compare with available observational estimates of thermally emitting isolated
neutron stars (INS)~\cite{Pot+20}. We recall that rapid cooling signals  large proton fractions, which render the direct Urca (DU) process possible at lower densities in comparison with softer models. Thus, rapid cooling signals a steep symmetry energy.

This paper is organized as follows. In sect.~\ref{eft}, we review our theoretical ingredients, omitting details that have been published elsewhere. In sect.~\ref{high}, we discuss continuations of the EoS above the microscopic predictions. In sect.~\ref{Cool}, we show preliminary predictions of cooling curves. A robust  analysis of neutron star cooling, including superfluid gaps and more, will appear in a later work.

\section{The equation of state at normal to moderately high density}
\label{eft}

\subsection{Theoretical framework} 
\label{TF} 
The theoretical framework we use to obtain the {\it ab initio} part of the equation of state has been published in detail elsewhere~\cite{SM21a, SM21b, SM22}, and thus we will not repeat  a lengthy presentation here. We will, however, briefly recall
 the spirit of chiral effective field theory (EFT), on which our nuclear forces are based.  A comprehensive and detailed review of our theoretical tools can be found in Ref.~\cite{Love}.

Given an energy scale and degrees of freedom appropriate at that scale, an EFT comprises all interactions consistent with the symmetries that govern
those degrees of freedom.
For the nuclear problem, 
relevant degrees of freedom are 
pions (Goldstone bosons), nucleons, and $\Delta(1232)$ isobars. We use the delta-less chiral EFT. To begin with, one writes the most general Lagrangians describing all interactions between pions, nucleons, and pions with nucleons. Because pion interactions must vanish at zero momentum transfer and in the chiral limit, where the pion mass,
$m_{\pi}$, goes to zero, the corresponding Lagrangian is expanded in powers of spatial derivatives
or pion masses. From these Lagrangians, an infinite number of Feynman diagrams can be generated, which seems to make the theory unmanageable. The strategy is then 
to design a scheme for ordering the
diagrams according to their importance -- the essence of 
Chiral Perturbation Theory (ChPT). Nuclear potentials are defined by the irreducible types among these
graphs.
(By definition, an irreducible graph is a diagram that
cannot be separated into two
by cutting only nucleon lines.)
These graphs are then analyzed in terms of powers of 
$Q$, with $Q=p/\Lambda_b$, 
where $p$ is generic for a momentum, (nucleon three-momentum
or pion four-momentum), or the pion mass, and $\Lambda_b \sim m_\rho \sim$ 0.7 GeV (with $m_{\rho}$ the mass of the $\rho$ meson) is the 
breakdown scale~\cite{Fur15}. Determining the power $\nu$ has become known
as power counting.
For a recent review of nuclear forces based on chiral EFT and their applications in nuclear and neutron matter, the reader is referred to Ref.~\cite{Love}.

 The neutron star crust is composed of metals in crystalline structure and cannot be described as a homogeneous fluid of nucleons, which is  an appropriate system for our microscopic approach. Instead, a crustal EoS~\cite{NV73} is joined to our previously described EoS {\it via} cubic spline interpolation. The crust is a Coulomb lattice of bound neutrons and protons clusters surrounded by a dilute neutron gas.

\paragraph{Quantifying errors in chiral EFT}
A reliable determination of the truncation error is a crucial aspect of chiral EFT.
 If observable $X$ has been calculated at order $\nu$ and at order $\nu+1$, a simple estimate of the truncation error at order $\nu$ is
\begin{equation}
\Delta X_{\nu} = |X_{\nu+1} - X_{\nu}| \; ,
\label{del} 
\end{equation} 
which is a measure for what is neglected at order $\nu$. A suitable prescription is needed
to estimate the uncertainty at the highest (included) order. For that purpose, we follow the prescription of Ref.~\cite{EKM15a}. If $p$ is of the order of the typical momentum involved in the system, the dimensionless parameter $Q$ is defined as the largest between $\frac{p}{\Lambda_{b}}$ and $\frac{m_{\pi}}{\Lambda_{b}}$, where $\Lambda_{b}$ is the breakdown scale, taken to be about 600 MeV. Before proceeding, some comments are in place to avoid confusion. In the pion-nucleon sector, it's natural to set the scale to the chiral symmetry breaking scale, $\Lambda_{\chi}$, about
 4$\pi F_{\pi} \approx$ 1 GeV, where $F_{\pi}$ is the pion decay constant, equal to 92 MeV (see Ref.~[21] of Ref.~\cite{EKM15a}). However, in the nucleon sector, it is common practice to apply a so-called breakdown scale, $\Lambda_b$, chosen around 600 MeV. This scale is smaller than $\Lambda_{\chi}$ because the non-perturbative resummation, necessary for nucleons, fails for momenta larger than approximately 600 MeV.

\begin{figure*}[!t] 
\centering
\hspace*{-1cm}
\includegraphics[width=8.7cm]{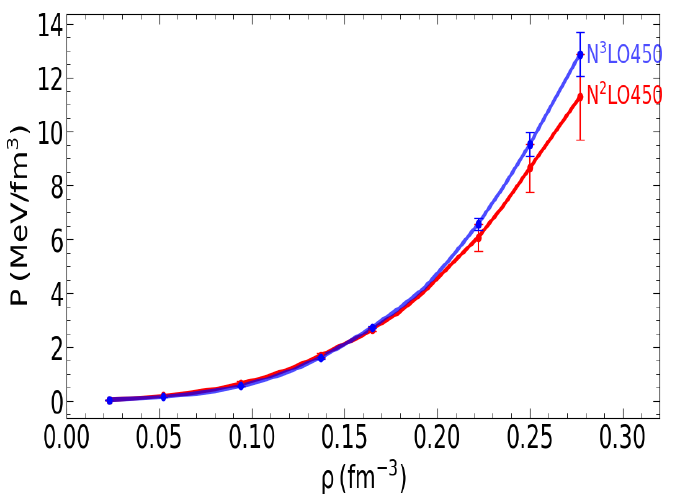}\hspace{0.01in}
\vspace*{-0.5cm}
\caption{Pressure as a function of density in $\beta$-stable matter at N$^2$LO (red) and at N$^3$LO (blue), with the respective truncation errors. In both cases, the predictions are based on the high-quality $NN$ potential of Ref.~\cite{EMN17} and include all 3NFs required at the respective order. }
\label{figpr}
\end{figure*}

Throughout the paper, we will show results at the (fully consistent) third order (N$^2$LO), and at the highest order which we have considered (fourth order, or N$^3$LO).
In Fig.~\ref{figpr}, we show the pressure as a function of density in $\beta$-stable matter at N$^2$LO (red) and at N$^3$LO (blue), with the respective truncation errors. In both cases, the predictions are based on the high-quality nucleon-nucleon ($NN$) potential of Ref.~\cite{EMN17} and include all three-nucleon forces (3NF) required at that order. For details on how our EoS are built, see, for instance, Refs.~\cite{SM21a,SM22}. They are available upon request from the corresponding author.

 The uncertainty in the value of observable $X$ at N$^3$LO as derived in Ref.~\cite{EKM15a} can be understood with the following arguments. If N$^3$LO ($\nu$ = 4) is the highest included order, the expression 
\begin{equation}
\Delta X_4 = |X_{4} - X_3| Q =( \Delta X_3) Q \; ,
\label{del43} 
\end{equation} 
is a reasonable estimate for $\Delta X_4$ in absence of the value $X_5$, because $Q$ to the power of 1 takes the error up by one order, the desired 4th order. To avoid accidental underestimations, a more robust prescription is to proceed in the same way for all the lower orders ($\nu$ = 0, 2, 3) and define, at N$^3$LO~\cite{EKM15a}:
\begin{displaymath}
\Delta X = \max \{Q^5|X_{LO}|, Q^3|X_{LO} - X_{NLO}|,Q^2|X_{NLO} - X_{N^2LO}|, 
\end{displaymath}
\begin{equation}
Q|X_{N^2LO} - X_{N^3LO}| \} \; .
\label{err}
\end{equation} 
In infinite matter, $p$ can be identified with the Fermi momentum at the density being considered.

 Cutoff variations have sometimes been used to estimate contributions beyond truncation. However, they do not allow to estimate the impact of neglected long-range contributions. Also, due to the intrinsic limitations of the EFT, a meaningful cutoff range is hard to estimate precisely, and often very limited. The method of Eq.~(\ref{err}) allows to determine truncation errors from predictions at all lower orders, without the need to use cutoff variations.

\paragraph{Chiral orders and three-nucleon forces: unresolved issues} 
\label{eos}

While the predictions at N$^2$LO are fully {\it ab initio}, 
a warning is in place for current N$^3$LO calculations.
As pointed out in Ref.~\cite{EKR20}, 
there is a problem with the regularized 3NF at N$^3$LO (and higher orders)
in all present nuclear structure calculations. 
The N$^3$LO 3NFs currently in use are regularized
by a multiplicative regulator applied to the 3NF expressions
derived from dimensional regularization.
This approach leads to violation of chiral symmetry at N$^3$LO
and destroys the consistency between two- and three-nucleon forces~\cite{EKR20, Ep+22}. Consequently, none of the current calculations that
include 3NFs at N$^3$LO (and beyond) can be considered truly {\it ab initio}.
An appropriate symmetry-preserving regulator~\cite{EKR20} should be applied to  
 the 3NF at N$^3$LO from Refs.~\cite{Ber08,Ber11}. 
At the present time, reliable predictions exist only at 
N$^2$LO, NLO, and LO. However, for the few fully {\it ab initio} calculations, the precision at N$^2$LO is unsatisfactory. A first step towards deriving consistently regularized nuclear
interactions in chiral EFT has been proposed in Refs.~\cite{KE23a,KE23b}. It requires the cutoff to be introduced already at the level of the effective Lagrangian. A path integral approach~\cite{KE23a}
can then be applied to the regularized chiral Lagrangian to derive nuclear forces through 
the standard power counting of chiral EFT. 

Throughout the paper, we will show results at the (fully consistent) third order (N$^2$LO), and at the highest order which we have considered (fourth order, or N$^3$LO).
In Fig.~\ref{figpr}, we show the pressure as a function of density in $\beta$-stable matter at N$^2$LO (red) and at N$^3$LO (blue), with the respective truncation errors. In both cases, the predictions are based on the high-quality nucleon-nucleon ($NN$) potential of Ref.~\cite{EMN17} and include all 3NFs required at that order. For details on how our EoS are built, see, for instance, Refs.~\cite{SM21a,SM22}. They are available upon request from the corresponding author.

\section{The equation of state at high density} 
\label{high}

It is important to emphasize that high-density EoS continuations are not meant to be a replacement for microscopic theories which, at this time, are not feasible in those regimes. Nevertheless, causality and maximum-mass constraints do pose considerable restrictions on the general features of the high-density EoS.

\begin{figure*}[!t] 
\centering
\hspace*{0.5cm}
\includegraphics[width=8.5cm]{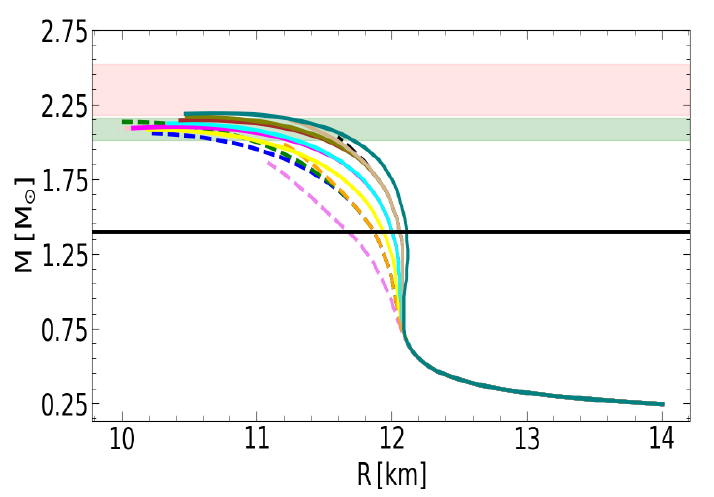}\hspace{0.01in}
\vspace*{-0.2cm}
 \caption{M(R) relations obtained with piecewise polytropes~\cite{SM22}. Equations of state that cannot support a maximum mass of at least 2.01 $M_{\odot}$ (see text) are discarded. The dashed curves are cut at the central density where causality is violated, whether or not the maximum mass has been reached. The black horizontal line marks the mass of the canonical neutron star, for reference. The green and pink shaded areas are constraints from J0740 + 6620~\cite{Fons+2021} and J0952 - 0.607~\cite{mmax}, respectively.
}
\label{ij}
\end{figure*}

Up to this point, we have used piecewise polytropes, which have the form:
\begin{equation}
P(\rho) = \alpha\Big (\frac{\rho}{\rho_0} \Big )^{\Gamma} \; ,
\label{pre}
\end{equation}
where $\rho_0$ is the density of saturated nuclear matter and $\Gamma$ is the adiabatic index. 
The corresponding energy density can be obtained from the basic relation between internal pressure and energy density,
\begin{equation}
P(\rho) =  \rho^2 \frac{d}{d\rho} \Big ( \frac{\epsilon}{\rho} \Big ) \; ,
\label{eps}
\end{equation}
or,
\begin{equation}
\epsilon(\rho) =\frac{ \alpha}{\Gamma - 1}  \Big (\frac{\rho}{\rho_0} \Big )^{\Gamma}  + c\rho \; .
\label{eps2}
\end{equation}
For a range of $\Gamma$ values, the parameters $\alpha$  and $c$ are determined by matching the values of $P$ and $\epsilon$ at the boundaries.
In the past, we accepted polytropes which can support a maximum mass of at least 2.01 $M_{\odot}$, to be consistent with the lower limit of the (2.08 $\pm$ 0.07) $M_{\odot}$ observation reported in Ref.~\cite{Mil21} for the J0740+6620 pulsar, along with a radius estimate of (\mbox{12.35 $\pm$ 0.75) km}.
 Figure~\ref{ij} displays results of the procedure we used in the recent past. The M(R) relations are obtained with piecewise combinations of two polytropes with different adiabatic index. Equations of state that cannot support a maximum mass of at least 2.01 $M_{\odot}$ (see above), are discarded, and solutions are cut at the central density where causality is violated~\cite{SM22}.  The initial range we considered for the adiabatic index, $\Gamma$, was approximately between 2.5 and 4.0, based on guidance from the literature, such as Ref.~\cite{Rea09}, where most of the EoS available from theory or phenomenology were fitted with polytropes.

 Currently, the maximum-mass constraint must account for the record-setting PSR J0952-0607, the heaviest neutron star found to date, at 2.35 $\pm$ 0.17 $M_{\odot}$~\cite{mmax}. We point out that this measurement was based on optical lightcurve modeling and may not be as accurate as those based on radio observations. For instance, for PSR J2215+5135, the optical lightcurve modeling suggests a mass of 2.27$\pm$ 0.17 $M_{\odot}$~\cite{Linares+2018}, while recent radio observations yield a significantly smaller value of 1.98 $\pm$ 0.08$M_{\odot}$~\cite{Sull+2024}. The analysis in Ref.~\cite{Fan+2024} found a maximum mass of 2.25 $\pm$ 0.07 $M_{\odot}$ for a non-rotating neutron star.

 We 
explored different piecewise parametrizations of the high-density EoS that preserve causality, while supporting masses at least as high as 2.2 $M_{\odot}$. We emphasize that {\it ab initio} predictions and most of the terrestrial constraints point to a soft symmetry energy at normal density, while the maximum mass constraint has moved to larger values. These considerations provide important guidance when building the phenomenological part of the EoS.

While checking different polytropic combinations, we made the observation that the ``best" combination (with regard to preserving causality while satisfying maximum-mass constraints) consists of a relatively stiff polytrope attached to the microscopic piece of the EoS, followed by a second, softer polytrope.

\begin{figure*}[!t] 
\centering
\hspace*{1.5cm}
\includegraphics[width=10.5cm]{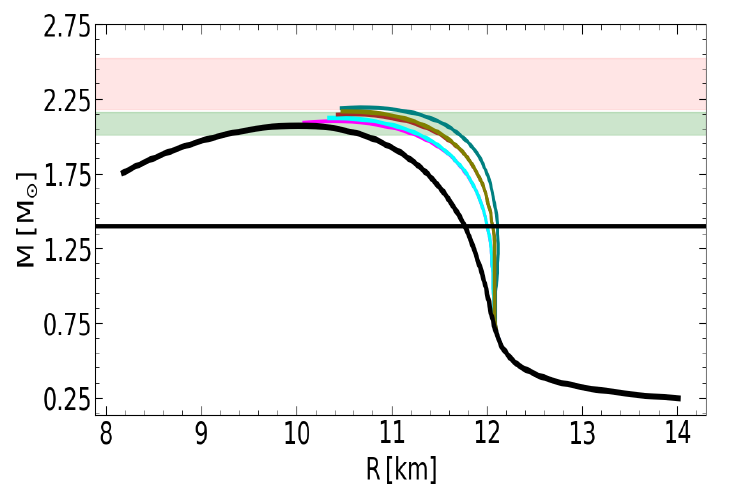}\hspace{0.01in}
\vspace*{-0.5cm}
 \caption{ Several M(R) relations. The curves in color are obtained from a sequence of two polytropes with adiabatic indices given in Table~\ref{gama}. The black curve is obtained with a single parametrization in terms of the speed of sound, as in Eq.~(\ref{pi}).
}
\label{ij+vs}
\end{figure*}

Although polytropic extension is a very general and popular method, alternative parametrizations of the high-density EoS offer desirable features~\cite{Kan+21, Tew18}, such as those in terms of the speed of sound.
 In Fig.~\ref{ij+vs}, the colorful curves are from selected EoS that generate maximum masses of about 2.1 to 2.2 solar masses and are consistent with causality. Table~\ref{gama} provides more information about these cases. We note that chiral uncertainties as those in Fig.~\ref{figpr} are not shown in the figures for the $M(R)$ relations. This is because chiral errors are meaningful only at the densities where the microscopic calculation is applied, which reach up to the central densities of the lighter stars (the "tail" of the $M(R)$ curves).
 The black curve is obtained with a single parametrization in terms of the speed of sound, constructed as follows~\cite{Kan+21,Tew18}.
Assigning $ i=0$ to values at threshold (the density at which the EoS parametrization has to be attached to the previous piece), we write
\begin{equation}
\rho_i = \rho_{i-1} + \Delta \rho \; ,
\label{eq4}
\end{equation}
\begin{equation}
\epsilon_i = \epsilon_{i-1} + \Delta \epsilon     \; ,
\label{eq5}
\end{equation}
and 
\begin{equation}
 \Delta \epsilon= \Delta \rho \frac{\epsilon_{i-1} + P_{i-1}}{\rho_{i-1}}  \; ,
\label{eq6}
\end{equation}
where we have used Eq.~(\ref{eps}).

The speed of sound is parametrized as 
\begin{equation}
\Big (\frac{v_s}{c} \Big )^2_i = 1 - c_1 exp \Big [ - \frac{(\rho_i - c_2)^2}{w^2}\Big ] \; , 
\label{vs}
\end{equation}
where $w$ is the width of the Gaussian curve, and the constants $c_1$ and $c_2$ are determined from continuity of the speed of sound and its derivative at the threshold density.
We note that the conformal limit, $\Big (\frac{v_s}{c} \Big )^2 \le 1/3$, is not imposed in Eq.~(\ref{vs}). 
Clearly, a larger value of $\Big (\frac{v_s}{c} \Big )^2 $ signifies increased pressure gradient to counterbalance the stronger gravitational force, and thus, larger masses. Therefore, observations of very massive stars require large values of the speed of sound at high density. A scenario where the speed of sound is not a monotone function of density would allow the conformal limit to be reached at asymptoticall large densities.

The pressure above the threshold is 
\begin{equation}
P_i = \Big ( \frac{v_s}{c} \Big )^2_{i-1} \Delta \epsilon + P_{i-1} \; .
\label{pi}
\end{equation}
This EoS continuation is manifestly causal at any density and reaches a maximum mass of 2.07 $M_{\odot}$.

\begin{table*}
\caption{Description of the M(R) relations in Fig.~\ref{ij+vs}.  }
\label{gama}
\centering
\begin{tabular*}{\textwidth}{@{\extracolsep{\fill}}ccccc}
\hline
\hline
 curve color  & $\Gamma_1$ &   $\Gamma_2$ & $M_{max}/M_{\odot}$ & $R_{1.4}$ (km) \\
\hline
\hline
 magenta & 3.1 & 2.7&  2.10 & 12.00   \\   
 cyan       & 3.1  & 2.8 & 2.12 &12.00 \\
brown     &  3.2  & 2.7 & 2.15 & 12.06 \\
olive        & 3.2  &  2.8 &2.17 & 12.06 \\
green     & 3.3 & 2.7 & 2.19 & 12.11 \\
\hline
\hline
\end{tabular*}
\end{table*} 

From the considerations above, we find that, for  the purpose of achieving high maximum masses while respecting causality at any density, a better solution is to combine a relatively steep (on the scale of Table~\ref{gama}) polytrope followed by a parametrization obtained from Eqs.~(\ref{vs}--\ref{pi}), which will maintain causality by construction. The matching densities are $\rho_1= 0.277 fm^{-3}$ and 
$\rho_2 = 0.563 fm^{-3}$.
The rationale for the first matching density is as follows. The {\it neutron} Fermi momentum in neutron matter, $k_F^n$,  at $\rho_1$ is equal to 2.02 $fm^{-1}$. Of course, this is larger than the momentum in beta-stable matter at the same density due to the presence of a proton fraction,
\begin{equation}
k_F^{snm} < k_F^{\beta} < k_F^n  \; ,
\label{kf}
\end{equation}
where $k_F^{snm}$ and  $ k_F^{\beta}$ are the Fermi momentum in symmetric nuclear matter and in beta-stable matter, respectively.
 The average momentum of a neutron Fermi gas is given by:
\begin{equation}
P_{av} = \sqrt{\frac{3}{5}} k_F^n  \; ,
\end{equation}
which we take as the typical momentum of the system, $p$, in defining the chiral expansion parameter, $ Q = \frac{p}{\Lambda_b}$, where $\Lambda_b$ was previously defined.  We obtain $Q = 51\%$, which is well below 1, and actually a pessimistic estimate, see Eq.~(\ref{kf}).
 For these reasons, we are comfortable applying the EFT up to this density.
The density $\rho_2$ is about two units of $\rho_0$ from the first matching point, a choice guided by Ref.~\cite{Rea09}, see section VB of that citation. We have tested the sensitivity of the M(R) results to moving this point out by one half of saturation density, and found it to be negligible. Moving that point toward lower densities brings down the maximum masses, which is not desirable. 

\begin{figure*}[!t] 
\centering
\hspace*{1.5cm}
\includegraphics[width=9.5cm]{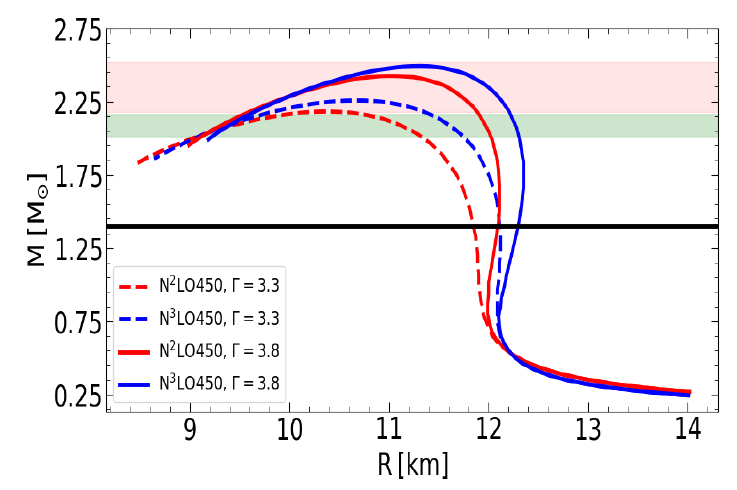}\hspace{0.01in}
\vspace*{-0.09cm}
 \caption{ M(R) curves at fourth order (N$^3$LO, blue) and at third order (N$^2$LO, red) of ChPT. Dashed curves: the first extension is done using a polytrope with $\Gamma$ = 3.3, followed by pressure values given by Eq.~(\ref{pi}) together with Eq.~(\ref{vs});  Solid curves: obtained with $\Gamma$ = 3.8, a value beyond which the EoS begins to violate causality. 
}
\label{7+vs}
\end{figure*}

\begin{figure*}[!t] 
\centering
\hspace*{1.5cm}
\includegraphics[width=7.5cm]{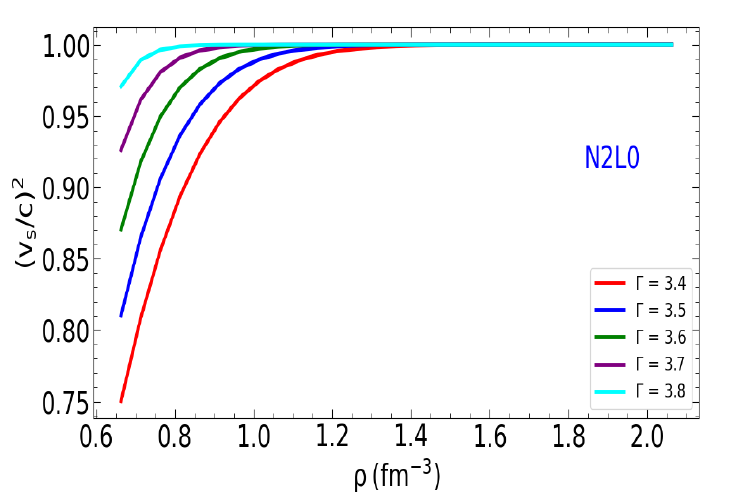}\hspace{0.01in}
\includegraphics[width=7.5cm]{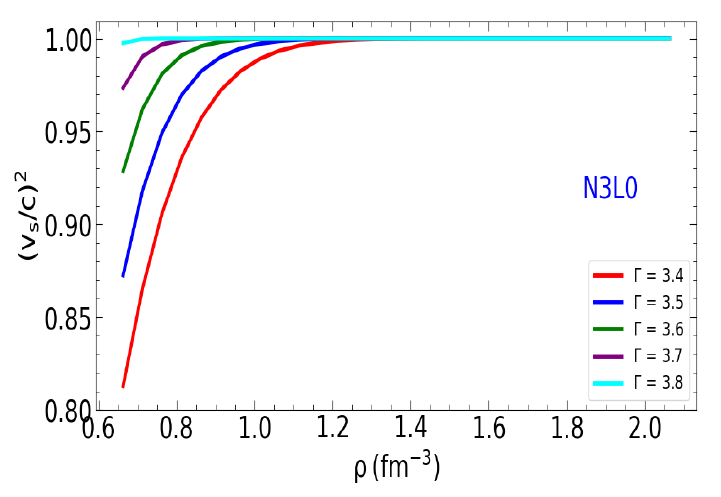}\hspace{0.01in}
\vspace*{-0.09cm}
 \caption{ Speed of sound, in units of the speed of light, at N$^2$LO (left) and at N$^3$LO (right). 
}
\label{vss}
\end{figure*}

The resulting $M(R)$ curves are shown in Fig.~\ref{7+vs} for both N$^3$LO (blue) and N$^2$LO (red). For the dashed curves, the first extension is done with a polytrope with $\Gamma$ = 3.3, followed by pressure values given by Eq.~(\ref{pi}) with the speed of sound (SoS) as in Eq.~(\ref{vs}). The solid curves (same color convention) have been obtained with $\Gamma$ = 3.8, a value beyond which the EoS begins to violate causality, see also Fig.~\ref{vss}. Table~\ref{vstable} displays the maximum mass,  its radius,  the central density,  and the radius of the canonical mass neutron star, for the curves in Fig.~\ref{7+vs}. We recall that the radius of a 1.4 $M_{\odot}$ is sensitive to the pressure at normal densities and thus it can pose constraints on microscopic theories of the EoS at those densities where such theories are applicable.

The speed of sound as a function of density is shown in Fig.~\ref{vss}. In each case, a single polytrope with the shown value of $ \Gamma$ is followed by the SoS-guided EoS. Thus, a polytrope that bridges the chiral EFT predictions with a causality-maintaining parametrization, has a limited range of powers. We underline that this scenario is inherently related to the softness of the chiral predictions. In other words, the nature of the predictions at normal density have a far-reaching impact, which extends to densities up to a few times normal density.
 \begin{table*}
\caption{ Some neutron star properties corresponding to the red and the blue M(R) relations shown in Fig.~\ref{7+vs}.}
\label{vstable}
\centering
\begin{tabular*}{\textwidth}{@{\extracolsep{\fill}}cccccc}
\hline
\hline
$ \Gamma$ & chiral order  & $M_{max}/M_{\odot}$ &   $R_{M_{max}} (km)$   & $\rho_c (fm^{-3}) $ & $R_{1.4}$ (km) \\
\hline
\hline
 3.3  & N$^2$LO & 2.18 & 10.34 &  1.12 & 11.84   \\   
        & N$^3$LO &  2.26     &  10.70   & 1.01   &  12.11   \\
3.8   & N$^2$LO & 2.42  & 10.99 &   0.94 & 12.09   \\   
        & N$^3$LO &  2.49    &  11.31   & 0.88   &  12.30   \\ 
\hline
\hline
\end{tabular*}
\end{table*} 

Of course, what we have presented is not the only option for building EoS that are consistent with current astronomical obervations.
We maintain, though, that an EoS must be ``bounded from below" by free-space few-nucleon data (which, in turn, have a strong impact on the symmetry energy and the pressure in neutron-rich matter at normal densities).
Typical examples of the other end of the spectrum are phenomenological EoS, such as those from Relativistic Mean Field (RMF) models. With no constraints from microscopic few-nucleon forces, new parametrizations can be constructed using
 different nonlinear, self- and 
inter-couplings among meson  and 
nucleon Dirac fields~\cite{Kum23}. Isovector mesons carry isospin dependence, with the main contribution to the symmetry energy coming from the pion~\cite{FS2011}.
In the RMF (pionless) framework,  the interplay between the isovector $\rho$ and $\delta$ mesons is described as the equivalent, in the isovector channel, of the $\sigma - \omega$ meson interplay in the isoscalar  channel. This approach, and the resulting couplings, have little to do with free-space $NN$ interactions~\cite{FS2011}. 
Not surprisingly, parametrizations can be found to cover a huge range of EoS ``stiffness," most recently incorporating CREX and/or PREX-II constraints~\cite{Kum23}.
Findings from RMF models concerning, especially, isovector quantities, such as the symmetry energy, must be interpreted with caution.

Before closing this section, we like to offer a few comments on EFT-inspired energy-density functionals (EDFs) that have been proposed in recent years.
Traditional EDFs are based on the mean-field (MF) approximation and built on empirical ingredients, which clearly prevent them from having truly predictive power. Furthermore, they do not follow 
a power-counting scheme, which prevents a reliable estimate of theoretical uncertainties. 
Attempts have been made to
render EDFs less empirical, trying to directly link them
to microscopic ingredients so as to reduce their intrinsic uncertainties~\cite{Grasso2019, BGY2020}. While this is potentially an improvement over traditional EDFs, it is our understanding that these schemes are perturbative in diluted matter. In fact, the YGLO (Yang-Grasso-Lacroix-Orsay)~\cite{Yang+2016} functionals are suitable at very low densities. Were they not perturbative at all, a connection with power counting would not be possible.

\section{Cooling of neutron stars}
\label{Cool}

\subsection{General considerations}

To create context, we review here some basic facts about  INS cooling.

Accurate modelling of neutron star cooling with account of all
possible effects is a complex problem. Cooling can be affected, for instance, by the presence of free hyperons or deconfined
quarks (see Ref.~\cite{Wei+2020} and references therein), and  pion or kaon
condensation (see Ref.~\cite{Yak01} and references therein). 

The internal structure of the neutron star can be taken, to a good approximation, to be spherically symmetric,
except for fast rotating INS or strong magnetic fields.
It is also reasonable to expect that the temperature distribution is
spherically symmetric at sufficiently high densities. Under these assumptions, the mechanical
structure and temperature distribution are determined by a set of
differential equations~\cite{Richard+1982} which involve only
one spatial coordinate, the radial coordinate $r$.

Neutron stars cool down mainly via neutrino emission
from their cores and photon emission from their atmospheres.
They are relativistic objects, and thus one needs to be careful about the coordinate system. The local temperature at some distance
 $r$ from the center is related to the temperature, $T^{\infty}$, measured by a distant observer, {\it via} the gravitational redshift between the coordinate systems:
\begin{equation}
T^{\infty} = e^{\phi(r)} T(r) \; ,
\end{equation}
where $\phi$ is the metric function.

The outermost layer of a neutron star is the {\it atmosphere},  consisting of gas
elements which emit thermal photons that can be observed on the Earth. Surface luminosity and temperature can be inferred by fitting this photon flux, and is a major source of cooling for
 older neutron stars. 
Below the atmosphere, there is a thin region called {\it envelope}, whose chemical composition is uncertain.

Although the distribution of the surface temperature over the
surface can be non-uniform, it is customary to approximate  the surface photon emission as the blackbody radiation from the entire surface. To that end, one introduces  the overall surface effective temperature of the star, $T_{s,eff}$, related to the photon luminosity, $L_{\gamma}$, by
\begin{equation}
\label{LvsT}
L_{\gamma} = 4 \pi \sigma_{SB} R^2 T_{s,eff}^4 \; ,
\end{equation}
where $\sigma_{SB}$ is the Stefan-Boltzmann constant.
The quantities in the above equation 
refer to a local reference frame at the neutron-star surface. Those 
 detected by a distant observer are redshifted,
\begin{equation}
L^{\infty}_{\gamma} = L_{\gamma} (1 - r_g/R) = 4 \pi \sigma_{SB} R^2_{\infty} (T_{s,eff}^{\infty})^4  \; ,
\end{equation}
\begin{equation}
T_{s,eff}^{\infty} = T_{s,eff} \sqrt{ 1 - r_g/R} \; , \; \; \; \; \; \; R_{\infty} = R /  \sqrt{ 1 - r_g/R} \; ,
\end{equation}
where $r_g$  is the Schwarzschild radius, $r_g = \frac{2GM}{c^2}$, with $G$ the gravitational constant.

 Either surface temperatures
or photon luminosities can be used to compare neutron-star observations with the
cooling theory. Both can be obtained with spectral analysis, but accurate determination is usually a challenge. 
One of the problems with obtaining accurate data suitable for testing
 the theory of cooling is that the vast majority
of neutron stars, including INS, emit intense radiation of non-thermal
origin. Neutron star binary systems are usually
surrounded by an accretion disk, whose emission is orders of
magnitude more powerful than the thermal emission from
the neutron star surface~\cite{Das+24}. Non-thermal emission of INS can also
be produced by other processes, and thus a careful analysis is required to
extract the thermal component of the observed spectrum. Another problem is that obtaining the ages of neutron stars from observation is difficult, and thus ages are only estimates.
Neutron stars that are estimated to be old have 
lost their initial heat, and therefore their thermal luminosity
is very low, and could have been 
produced by reheating~\cite{Gonz+2015, Gonz+2019}.

 In summary, the ``standard" cooling
theory, which neglects reheating, can only be tested against 
observations of a small fraction of INS, using estimated ages.

\subsection{Proof of concept results}
\label{res}

In this section, we perform cooling simulations employing the two EoS used to generate the red and blue $M(R)$ dashed curves in Fig.~\ref{7+vs}.  Our beta-stable EoS include protons, electrons, and muons. We emphasize  that these are preliminary curves, to have a first look at the mass dependence in relation to the available data. In other words, these are proof of concept calculations, to build upon.

\begin{figure*}[!t] 
\centering
\hspace*{0.9cm}
\includegraphics[width=7.8cm]{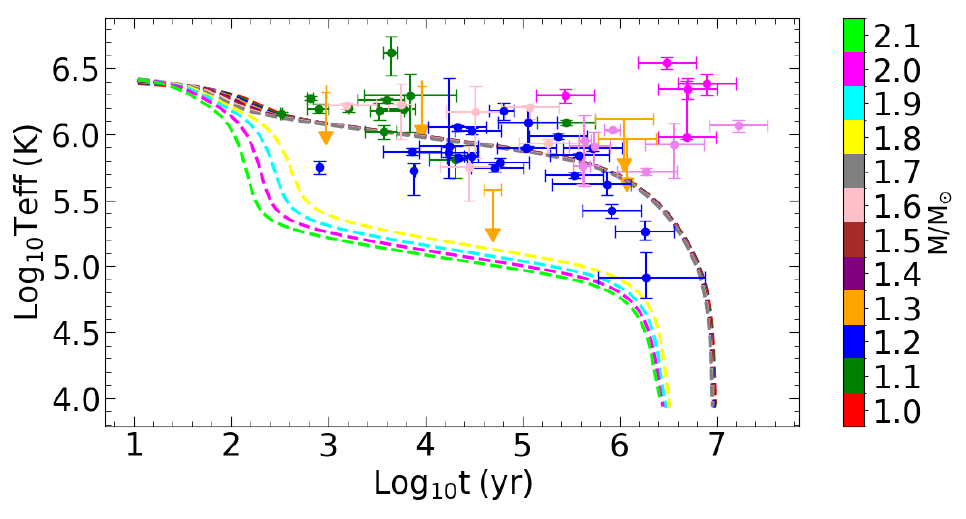}\hspace{0.01in}
\includegraphics[width=7.8cm]{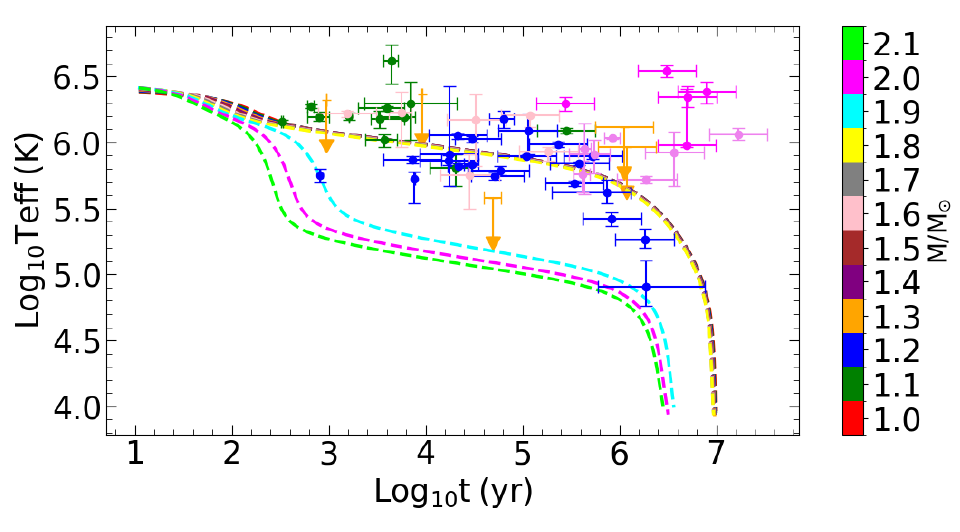}\hspace{0.01in}
\vspace*{0.1cm}
\caption{Effective temperature as a function of time, for different masses (colors). Left: N$^2$LO; Right: N$^3$LO. The envelope from Ref.~\cite{Pot+1997} is applied. } 
\label{figtemp}
\end{figure*}

\begin{figure*}[!t] 
\centering
\hspace*{0.9cm}
\includegraphics[width=7.8cm]{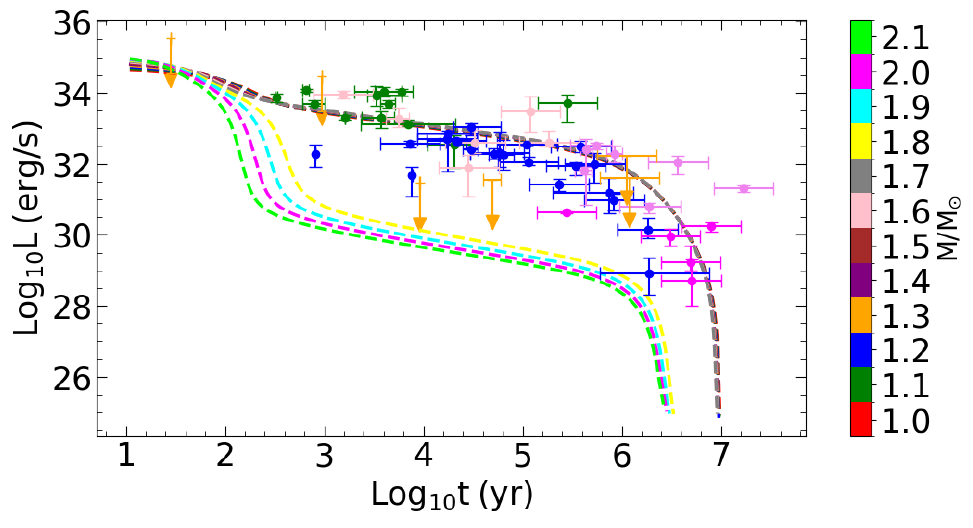}\hspace{0.01in}
\includegraphics[width=7.8cm]{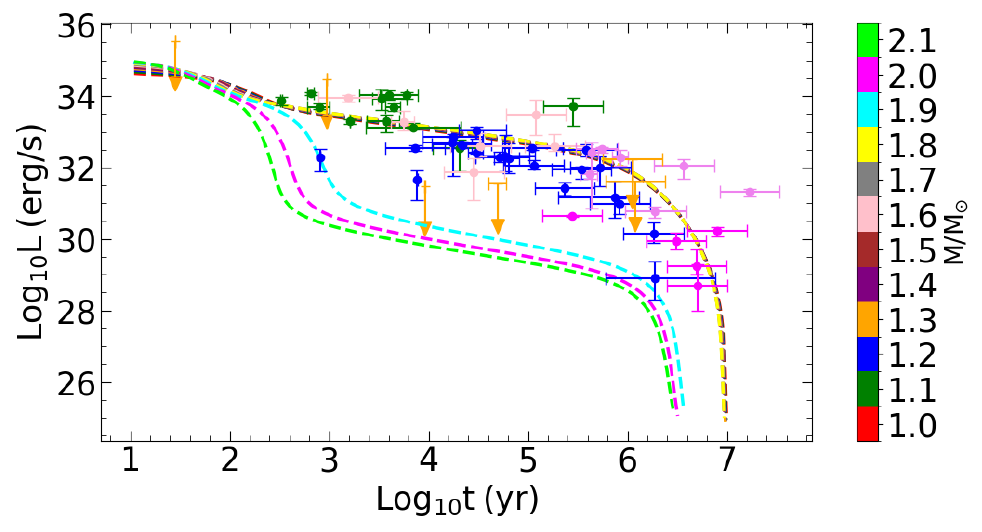}\hspace{0.01in}
\vspace*{0.1cm}
\caption{Photon luminosity as a function of time, for different masses (colors). Left: N$^2$LO; Right: N$^3$LO. All conditions as in Fig.~\ref{figtemp}.}
\label{figlum}
\end{figure*}

From Figs.~\ref{figtemp} and ~\ref{figlum}, 
one can see the mass dependence of the effective temperature and the closely related luminosity, see Eq.~(\ref{LvsT}). The more massive INS correspond to faster cooling, suggesting that enhanced neutrino emission
due to DU reactions operates in those stars, where the proton fraction
in the interior reaches values sufficient to enable the process.
Pairing, not included here, could suppress DU processes. The data are from Ref.~\cite{pot+}. 

\begin{figure*}[!t] 
\centering
\hspace*{0.9cm}
\includegraphics[width=7.8cm]{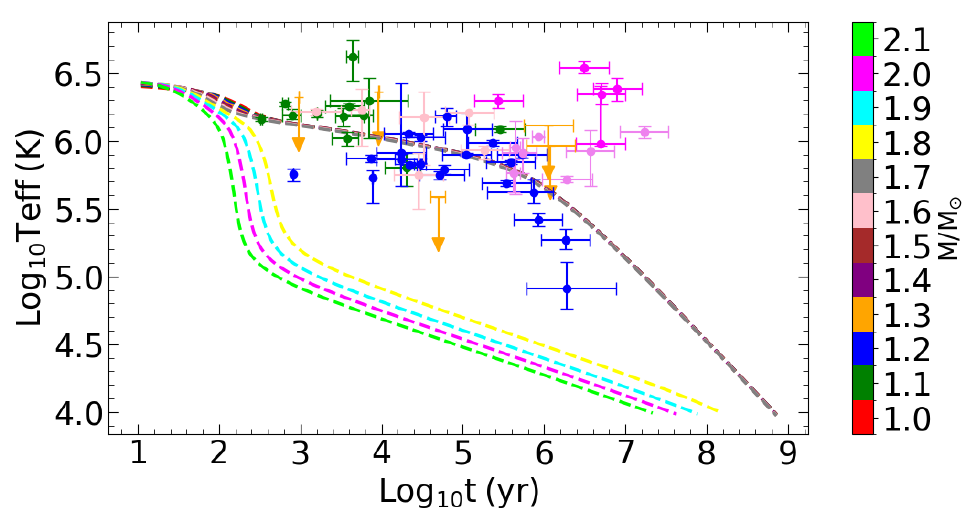}\hspace{0.01in}
\includegraphics[width=7.8cm]{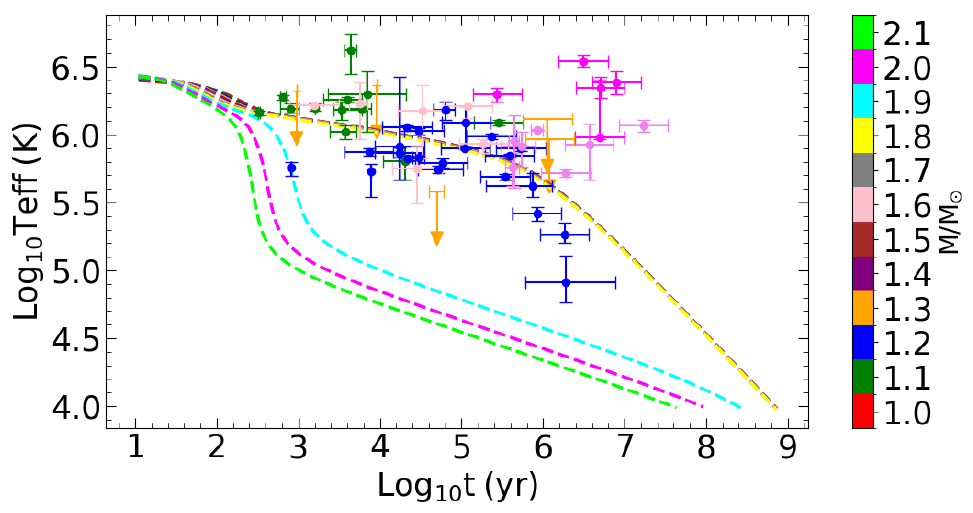}\hspace{0.01in}
\vspace*{0.1cm}
\caption{As in Fig.~\ref{figtemp}, but with envelope model from Ref.~\cite{Nom+1987}.}
\label{figtemp_2}
\end{figure*}

The difference between Fig.~\ref{figtemp} and Fig.~\ref{figtemp_2} is the model for the envelope. In Fig.~\ref{figtemp}, the envelope contains light elements up to densities where they can still be present, and heavier elements, including iron, at the higher densities~\cite{Pot+1997}. More precisely, blanketing envelopes are composed, from surface to bottom, of hydrogen, helium, carbon, and iron shells, in a stratified structure. In Fig.~\ref{figtemp_2}, older iron models for the envelope~\cite{Nom+1987} are employed to gauge the sensitivity to the chemical composition of the envelope. We find the latter to have a significant impact on the cooling curves, especially for low to medium mass neutron stars.
The envelope acts as a thermal insulator between
the surface and the hot interior, thus relating interior temperature to the star's effective surface temperature.
 There is a large temperature gradient between the
top and bottom layers of the envelope, determined by the amount of light
elements such as Hydrogen or Helium. Therefore, the composition of the envelope  impacts its photon cooling. 

Some investigations~\cite{Das+24} have concluded that an EoS  allowing DU cooling for a wide
enough mass range of neutron stars, combined with some quenching
by the proton $^1S_0$ BCS gap, agrees best with the
cooling data, while the neutron pairing gap in the triplet $P$-wave 
seems to generate overly rapid cooling. Others~\cite{KPW23} find that pairing in the triplet $P$-states prevail in neutron matter, but essentially disappear if the spin-orbit interaction is turned off. Overall, the contribution from pairing is reported to be quite sensitive to the characteristics of the model. In fact, the sensitivity of the gap to the input interaction and medium effects can be dramatic, which calls for 
 further investigations. We will introduce short-range correlations (SRC) by replacing in the gap equation the bare potential with the G-matrix. The latter will be calculated self-consistently with the single-particle potential, which we will use in the single-particle spectrum. Typically, one would expect SRC to reduce the gap by introducing more high-momentum components and thus removing strength around the Fermi level and depleting the gap~\cite{RPD17}. 
As SRC are the most model-dependent part of an interaction, they certainly contribute to the gap's model dependence.
At the same time, availability of more and more accurate data from INS is crucial to constrain all important aspects of the theoretical input.

\section{Conclusions and work in progress}

The intrinsic and strong relation between the EoS and the maximum mass of a neutron star sequence is a remarkable feature. In fact, knowledge of one is essential to access the other.
In our observations,  the maximum-mass constraint moving to higher values, together with the causality requirement at any central density, poses significant restrictions on the high-density EoS.  The softness of the microscopic predictions at normal density brings up the need for a (first) steeper extension. A scenario such as the one we have described, where the first part of a piecewise extension needs to become stiffer in order to support current maximum mass constraints, while the next piece must soften to maintain causality, would suggest that phase transitions and/or new species (which can occupy their lowest energy states), begin to appear only at the highest densities, where the composition of a neutron star is likely to remain unknown in the foreseeable future. Also, transport properties of superdense matter where the speed of sound approaches the speed of light are essentially unknown.

A monotone behavior of the speed of sound approaching the conformal limit seems to be excluded by mass constraints, which require a rapid growth to allow masses of 2 $M_{\odot}$ or above.  A scenario where the speed of sound peaks around few to several times nuclear density and then falls back to approach the QCD limit for deconfined quark matter would signify some sort of phase transition with the conformal limit reached well beyond central densities of the heaviest observed neutron stars. The superconformality condition satisfied on the average in neutron stars, $< \Big (\frac{v_s}{c} \Big )^2  > 1/3$, would have fundamental implications on the trace of the energy-momentum tensor at the center of rotating neutron stars~\cite{Raissa+24}.

We reiterate that a microscopic theory of the nuclear many-body problem must start from quantitative descriptions of few-nucleon interactions. Those constraints have implications at normal density and well beyond it.

We also took the opportunity to display cooling curves as the foundation of a forthcoming comprehensive analysis, including gaps and medium effects. 
Based on the available literature, one may conclude that the impact of including 3NFs or other medium effects 
in calculations of the triplet pairing gap in neutron matter vary wildly, both quantitatively and qualitatively,
depending on the specifics of the input. Systematic studies with robust two- and three-nucleon forces are called for.

\section*{Acknowledgments}
This work was supported by 
the U.S. Department of Energy, Office of Science, Office of Basic Energy Sciences, under Award Number DE-FG02-03ER41270. \\
The authors are grateful to Dany Page for help with the NSCool cooling simulation package.

\bibliography{bibRM}
\end{document}